\documentclass[runningheads]{llncs}
\usepackage{graphicx}

\usepackage[cmex10]{amsmath}
\usepackage{amssymb}
\usepackage{amsmath}
\usepackage{tabularx}
\newcolumntype{C}{>{\centering}X}
\usepackage{filecontents,lipsum}
\usepackage[noadjust]{cite}
\usepackage{graphicx}
\usepackage{caption}

\begin{document}
\title{Secure Multilayer Perceptron Based On Homomorphic Encryption}
%
%
\author{Reda Bellafqira\inst{1}\orcidID{2} \and
Gouenou Coatrieux\inst{1}\orcidID{2} \and
Emmanuelle Genin\inst{2}\and
 Michel Cozic\inst{3}}
\authorrunning{R. Bellafqira et al.}
%
\institute{IMT Atlantique, 655 Avenue du Technopole, 29200 Plouzane, France\\
\email{\{reda.bellafqira, gouenou.coatrieux\}@imt-atlantique.com}  \and
Unité INSERM 1101 Latim, 29238 Brest Cedex, France \\
  \email{ emmanuelle.genin@inserm.fr}
\and MED.e.COM, Plougastel Daoulas 29470, France  \\
\email{mcozic@wanadoo.fr }\\
 }

\maketitle              
\begin{abstract}
In this work, we propose an outsourced Secure Multilayer Perceptron (SMLP) scheme where privacy and confidentiality of both the data and the model are ensured during the training and the classification phases. More clearly, this SMLP : i) can be trained by a cloud server based on data previously outsourced by a user in an homomorphically encrypted form; ii) its parameters are homomorphically encrypted giving thus no clues to the cloud; and iii) it can also be used for classifying new encrypted data sent by the user returning him the encrypted classification result encrypted. The originality of this scheme is threefold. To the best of our knowledge, it is the first multilayer perceptron (MLP) secured in its training phase over homomorphically encrypted data with no problem of convergence. And It does not require extra-communications between the server and the user. It is based on the Rectified Linear Unit (ReLU) activation function that we secure with no approximation contrarily to actual SMLP solutions. To do so, we take advantage of two semi-honest non-colluding servers. Experimental results carried out on a binary database encrypted with the Paillier cryptosystem demonstrate the overall performance of our scheme and its convergence.
\end{abstract}
\section{Introduction}
\label{section1}
Nowadays, cloud technology allows outsourcing the processing and/or storage of huge volume of data, these ones being personal data or data issued from many sources for big-data analysis purposes. In healthcare domain, for example, different initiatives aim at sharing medical images and Personal Health Records (PHR) between health professionals or hospitals with the help of cloud \cite{decenciere2013teleophta}. They take advantage of the medical knowledge this volume of data represents so as to develop new decision making tools based on machine learning techniques. Among such techniques, there is the multilayer perceptron (MLP) method which belongs to Neural Network (NN) family and which is a core element of deep learning methods; methods that are broadly studied and used nowadays. A MLP consists of multiple layers of interconnected perceptrons (see Fig. \ref{Fig3}). A perceptron is a classifier that maps its inputs with a vector of weights followed by an activation function. The output of a perceptron is the input of the next perceptron layer. As all machine learning algorithms, MLP works in two distinct ways: the training phase and the classification of new data. In a supervised mode, the training phase aims at inferring the network parameters from a labeled database by optimizing some objective function. Once trained, a MLP scheme is used so as  to classify new data.
 
Despite the attractive benefits provided by MLP, one of the actual limits of its outsourcing in a cloud environment stands on the security of data used for the training phase or for classification purposes issues as well as of the MLP parameters. Indeed, externalization involves that the user looses the control on his data and on their processing \cite{chow2009controlling}. Recent news show a clear evidence that outsourced storage is not safe against privacy threats, these ones being external (e.g., hackers \cite{lewis2014icloud}) or internal \cite{chow2009controlling,WinNT}. At the same time, the parameter of a process, like those of a trained MLP, may have some important added value for a company, for example. There is thus an interest to develop secured MLP (SMLP) methods that can be trained remotely using outsourced data while respecting data privacy and confidentiality.

Different approaches have been proposed to secure neural network methods. Some of them are based on additive secret sharing that allows several parties to jointly compute the value of a target function $f(.)$ without compromising the privacy of its input data. For instance, \cite{39schlitter2008protocol} present a privacy preserving NN learning protocol where each user performs most of the learning computations in the clear domain except the NN weight update which is performed using the secret sharing. One limit of this solution is that updated NN weight values are revealed to all users at each iteration, giving thus some clues about users' training data. To reduce such information leakage, \cite{42shokri2015privacy} proposes to only share a small fraction of parameters which the NN weights update can be performed. This method consequently establishes a compromise between accuracy and privacy. Higher the number of shared parameters, better is the classification accuracy but lower is privacy. To go further, and in addition to the secure NN weight update, \cite{12chen2009privacy} aims at securing the NN activation function with additive secret sharing. Because activation functions are usually non-linear (e.g. most MLP work with the Sigmoid and ReLU activation functions), the authors of \cite{12chen2009privacy} linearly approximate them, introducing at the same time convergence issues in the MLP training phase. An alternative to these approaches is to train the neural network in clear domain and by next use it with encrypted data. Most solutions make use of homomorphic encryption the interest of which is that it allows performing operations (e.g. $+$ , $\times$) onto encrypted data with the guarantee that the decrypted result equals the one carried out with unencrypted data \cite{bellafqira2017proxy, bellafqira2016end, bellafqira2015content}. To the best of our knowledge, the NN training phase with homomorphically encrypted data has only been theoretically studied in \cite{xie2014crypto}. It is shown that NN can be trained using fully homomorphic encryption data  and by approximating the activation functions with polynomials. However, with fully homomorphic cryptosystems, both the computational complexity and the length of cipher-texts increase with the number of desired operations in order to guarantee the correct decryption after polynomial evaluation. While these increase slowly by increasing the number of additions, it is more important when adding multiplications. Therefore, practical implementations should have a denoising process in order to be feasible or to restrict the computation just on low degree polynomials. All other proposals focus on the NN classification phase. As example, \cite{bost2015machine} proposes three privacy homomorphic encryption based on three classifiers: the linear and two low degree models. In \cite{chabanne2017privacy} a fully homomorphic convolutional neural networks classifier (CNN) is proposed. 

In this paper, we propose a secure multilayer perceptron (SMLP) method the training and classification procedures of which do not suffer of convergence issues. To do so, we take advantage of the rectified linear unit (ReLU). Beyond its accuracy and its contribution to MLP efficiency \cite{Reluglorot2011deep}, ReLU can be secured with homomorphic encryption and two non-colluding semi-honest servers avoiding thus the need to use an approximation procedure of the perceptron's output as proposed by the above methods. Another originality of our SMLP, is that its output is also encrypted. That is not the case of actual solutions that provide unencrypted output.  Furthermore, our SMLP is entirely outsourced in the sense that it does not require extra-communication overhead in-between the servers and the user to conduct the training and classification phases. The user just has to send his data homomorphically encrypted to the cloud server that will train the SMLP or classify data, without the cloud being able to infer information about the SMLP model parameters, the data or the classification result.

The rest of this paper is organized as follows. Section~\ref{section2} regroups preliminaries related to Multilayer Perceptron and the Paillier cryptosystem on which relies the implementation of our SMLP. We also provide the basic properties and the operations one can implement over Paillier encrypted data when using two non-colluding semi-honest servers. In Section~\ref{section3}, we detail our secure multilayer perceptron. Section~\ref{section4} provides some experimental results conducted to model the "AND" logic function on a binary database, and the security analysis of our proposal. Section~\ref{section5} concludes this paper. 
\begin{figure}[t!]
\begin{center}
\includegraphics[scale=.5]{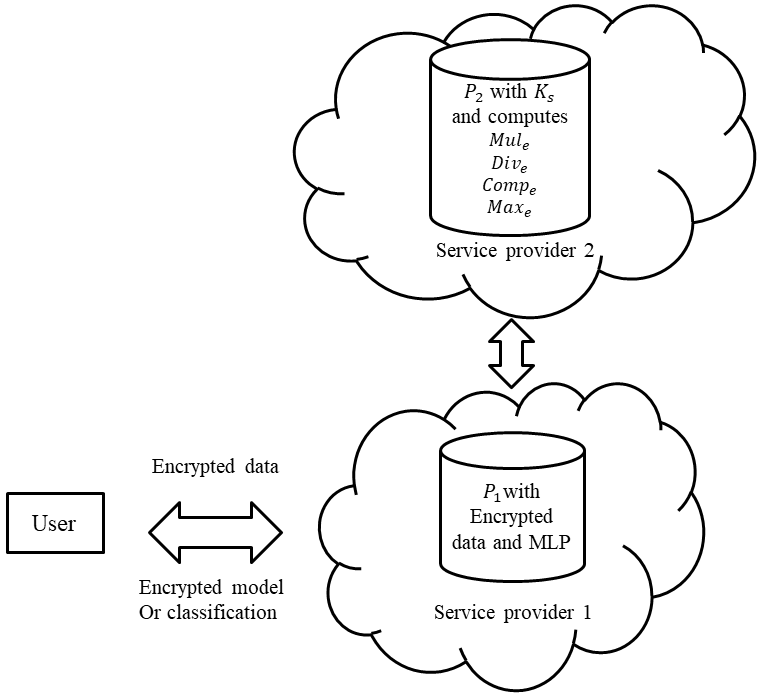}
\end{center}
\caption{\label{Fig1}Secure neural network architecture in a cloud environment. }
\end{figure}
\section{Preliminaries on the Paillier cryptosystem and Multilayer Perceptron}
\label{section2}
\subsection{The Paillier cryptosystem}
\label{section2-1}
Being asymmetric, the Paillier cryptosystem \cite{paillier1999public} makes use of a pair of private and public keys. Let $p$ and $q$ be two large primes. Let also $K_p=pq$, $\mathbb{Z}_{K_p}= \{0, 1,..., K_p-1\}$,  $\mathbb{Z}_{K_p}^*$ denotes the set of integers that have multiplicative inverses modulo $K_p$ in $\mathbb{Z}_{K_p}$. We select also $g\in\mathbb{Z}_{K_p^2}^*$ such as
\begin{equation}
	gcd(L(g^\lambda \mod K_p^2),K_p)=1
\label{(1)}
\end{equation}
where: $gcd(.)$ is the greatest common divisor function; $\lambda=lcm(p-1,q-1)$ is the private key ($K_s$), with $lcm(.)$ the least common multiple function; the pair $(K_p,g)$ defines the public key; and, $L(s)=\frac{s-1}{K_p}$. 
Let $m\in \mathbb{Z}_{K_p}$ the message to be encrypted. Its cipher-text is derived as
\begin{equation}
        c= E[m,r]= g^m r^{K_p} \mod K_p^2
        \label{(2)}
\end{equation}
where $E[.]$ is the encryption function, $r \in \mathbb{Z}_{K_p}^*$ is a random integer. Since $r$ is not fixed, the Paillier cryptosystem satisfies the so-called "semantic security". More  clearly, depending on the value of $r$, the encryption of the same plain-text message yields to different cipher-texts even though the public encryption key is the same. The plain-text $m$ is recovered using the decryption function $D[.]$ defined as follow
\begin{equation}
m=D[c,\lambda]= \frac{L(c^\lambda  mod K_p^2 )}{L(g^\lambda  mod K_p^2 )} \mod K_p
\label{(3)}
\end{equation}
The Paillier cryptosystem has an additive homomorphic property. Considering two plain-texts $m_1$ and $m_2$, then
\begin{equation}
E[m_1,r_1].E[m_2,r_2] = E[m_1+m_2, r_1.r_2]
\label{(4)}
\end{equation}
\begin{equation}
E[m_1,r_1]^{m_2} = E[m_1.m_2, r_1^{m_2}]
\label{(5)}
\end{equation}
For the sake of simplicity, in the sequel we denote $E[m,r]$ by $E[m]$.
\subsection{Operations over Paillier encrypted data}
\label{section2-2}
As stated above, the Paillier cryptosystem allows implementing linear operations. It can however be used to compute multiplications, divisions and comparisons with the help of two non-colluding semi-honest servers $P_1$ and $P_2$. 
\begin{itemize}
\item Multiplication operator in Paillier encrypted domain $Mul_e^{P_1, P_2} (.;.)$

Let us consider two messages $a$ and $b$ and their respective Paillier encrypted versions $E[a]$ and $E[b]$ obtained with the user public key $K_p$. In order to compute $E[a\times b]$ without revealing any information about $a$ and $b$, one can take advantage of blinding and two servers $P_1$ and $P_2$. Assuming that $P_1$ possesses $(E[a],E[b])$, the objective is that $P_2$ returns $E[a\times b]$ to $P_1$ while ensuring that no clues about $a$ and $b$ are revealed to $P_1$ and $P_2$. Under the hypothesis $P_2$ knows the user secret key $K_s$ and that it does not collude with $P_1$, this objective can be reach according to the following procedure we will refer as $Mul_e^{P_1, P_2} (a;b)$: 
\begin{enumerate}
\item \textit{Data randomization} - $P_1$ firstly randomizes $E[a]$ and $E[b]$ such that: 
\begin{align}
a'=E[a]\times E[r_a ]=E[a+r_a]\\
b'=E[b]\times E[r_b ]=E[b+r_b]
\label{(6)(7)}
\end{align}
where $r_a$ and $r_b$ are two random numbers only known from $P_1$ and uniformly chosen in $\mathbb{Z}_{K_p}$. Then $P_1$ sends $a'$ and $b'$ to $P_2$.
\item \textit{Multiplication computation phase} - On its side, using the user private key $K_s$, $P_2$ decrypts $a'$ and $b'$ and multiplies the result
\begin{equation}
M=(a+r_a )(b+r_b )
\label{(8)}
\end{equation}
$P_2$ next encrypts $M$ into $E[M]$ using the user public key $K_p$ and sends it to $P_1$.
\item 	\textit{Multiplication denoising} - In order to get $E[a\times b]$, $P_1$ just has to remove the extra-random factors as follow 
\end{enumerate}
\begin{equation}
E[a\times b]=E[M]\times E[b]^{-r_a}\times E[a]^{-r_b}\times E[-r_a\times r_b ] 
\label{(9)}
\end{equation}	
\item 	Division operator in Paillier encrypted domain: $Div_e^{P_1, P_2} (.;.)$

Different ways, based on two servers, have been proposed so as to compute the division . The one used in this paper works as follows. Let us consider $P_1$ has an encrypted message $E[a]$ and that it wants to divide $a$ by $d$. At this time $d$ can be encrypted or not, that is to say known or unknown from $P_1$. Again, we don't want $P_1$ and $P_2$ to learn details about $a$. The computation of $E[a/d]$ from $E[a]$ and $d$ is also based on blinding. As above, it is assumed that $P_2$ possesses the decryption key $K_s$. Our division operation $Div_e^{P_1, P_2} (a;d)$ is thus a procedure defined as:
\begin{enumerate}
 \item \textit{Data blinding} -  $P_1$ randomly chooses a number $r \in \mathbb{Z}_{K_p}$ and computes $E[z]=E[a+r]=E[a]E[r]$. $P_1$ then sends $E[z]$ to $P_2$. 
 \item \textit{Division computation} - $P_2$ decrypts $E[z]$ with the user private key $K_s$ and computes $c=z/d$. $P_2$ encrypts the division result $E[c]$ and sends it to $P_1$.
 \item  \textit{Division denoising} - $P_1$ computes $E[a/d]$ such as:
\begin{equation}
	E[a/d]=E[c]\times E[-r/d].  	 
\end{equation}	 
\end{enumerate}
\end{itemize}	
\subsection{Multilayer Perceptron}	
\label{section2-3}
\begin{figure}[t!]
\begin{center}
\includegraphics[scale=.5]{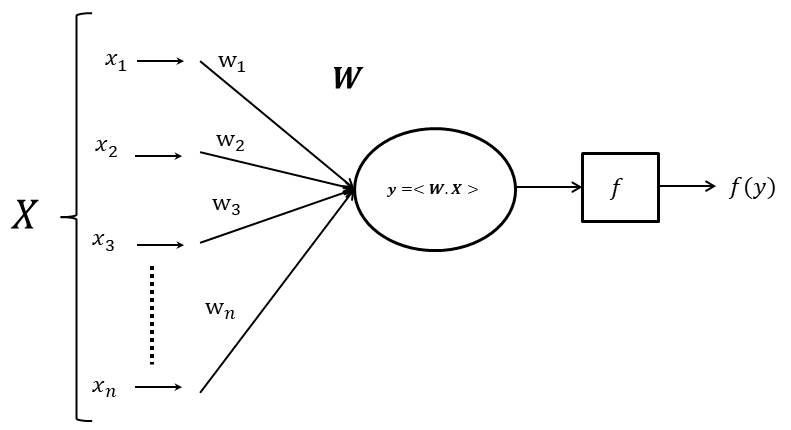}
\end{center}
\caption{\label{Fig2}Architecture of a single Perceptron. }
\end{figure}

The common architecture of a MLP is given in Fig. \ref{Fig3}. It is constituted of perceptrons organized in different layers: the input and the output layers and, in-between them, a given number of hidden layers (one in the given example of Fig. \ref{Fig3}). The first layer takes as input the user data as a vector $X=\{x_i\}_{i=1...n}$ where $n$ is the number of perceptron inputs, while the output layer provides the class of the input data. As illustrated in Fig. \ref{Fig2}, a perceptron is a classifier that maps its input (an integer value vector $X$) to an output $z$ value: 
\begin{equation}
	z=f(y=<W.X>)=f( \sum_{i=1} w_i.x_i ) = f(y)
	\label{weiSum}
\end{equation}
where $W$ is a weight vector of size $n$ and $f()$ an activation function. Many activation functions have been proposed (e.g. Sigmoid, Tanh, ReLU ). In this work, we opted for the rectified linear unit (ReLU) activation function, one of the most used function due to its accuracy and its efficiency \cite{Reluglorot2011deep}. Another reason is that it can be secured by means of homomorphic operators. We will come back in details on this point in Section \ref{section3}. ReLU is defined such as 
\begin{equation}
f(y)=\left\lbrace \begin{array}{ccc}
       y & $if$ & y\ge 0 \\
       0 & $otherwise$  	  	
	\end{array} \right.
\label{relu}
\end{equation}
To make such a MLP scheme operational, it should be trained so as to find the perceptron weigth values. This training can be supervised or unsupervised. In the former case, the classes of data the MLP should distinguish are \textit{a priori} known. Thus, to train a MLP scheme, the user provides labeled data $T=\{t_i \}_{i=1...K}$ where $K$ is the size of the training set and $t_i$ indicates the class label of the $i^{th}$ training input data. In the second case, the perceptron will identify by itself the different classes of data. The solution we proposed in this work is trained on labeled data.

The supervised training of NN relies on two phases that are iteratively applied: the feed-forward and the back-propagation phases. Before starting the first feed-forward phase, all perceptrons' weights are initialized with random values, for instance. Then training data are provided as input to the MLP. By next, the error between the original label and the ones computed by the perceptron is calculated using an objective function (called also cost function) (e.g. cross entropy, Mean Square Error, Minkowski distance). This error is then used in the back-propagation phase, so as to update all perceptrons' weights applying gradient descent. Once weights updated, a new feed-forward starts using the same labeled data. 

Many solutions have been proposed so as to decide when to stop the training phase  \cite{castellano2000variable}.  Among these conditions, one can fix a number of iterations ("aka epochs"): the MLP will stop once a number of epochs have elapsed. An alternative stands in thresholding the training set Mean Squared Error (MSE) (i.e. MSE between the training set labels and the MLP outputs). The smaller MSE, the network better  performs. Thus the training stops when MSE is smaller than a given threshold value. Instead, it has been proposed to use the Training Set Accuracy; that is the number of correctly classified data over the total number of data in the training set for an epoch. In this work, we opted for a fix number of iterations.

Once a MLP model or scheme trained, i.e. once the perceptrons' weights known, it can be used in order to classify new data. This classification process simply consists in applying the feedforward phase with new data as input, considering that the MLP output will give the data class.  

\section{Secure Multilayer Perceptron }
\label{section3}
\subsection{General framework and system architecture}
\label{section3-1}
The general framework we consider in this work is given in Fig. \ref{Fig1}, where a user outsourced into the cloud Paillier encrypted data; data on which the user wants  the cloud service provider to train our secure multilayer perceptron (SMLP). Once trained, this SMLP will be used by the user so as to classify new data. He will also send encrypted to the cloud. In our view, the data, the classification data result as well as all the parameters of the SMLP should be unknown from the cloud. As it can be seen in Fig. \ref{Fig1} and as we will see in the sequel, the computations of both the SMLP training and classification phases are distributed over two servers, $P_1$ and $P_2$, of two distinct cloud service providers. We consider them as honest but curious \cite{huang2011faster}. More clearly, they will try to infer information about the data, the classification results as well as about the SMLP parameters. In our scenario, $P_1$ interacts with the user and stores and handles his data. $P_2$ cooperates with $P_1$ so as to conduct some operations (division, multiplication, etc ...) involved into the training or classification phases of ours MLP.
 \begin{figure}[h]
\begin{center}
\includegraphics[scale=.5]{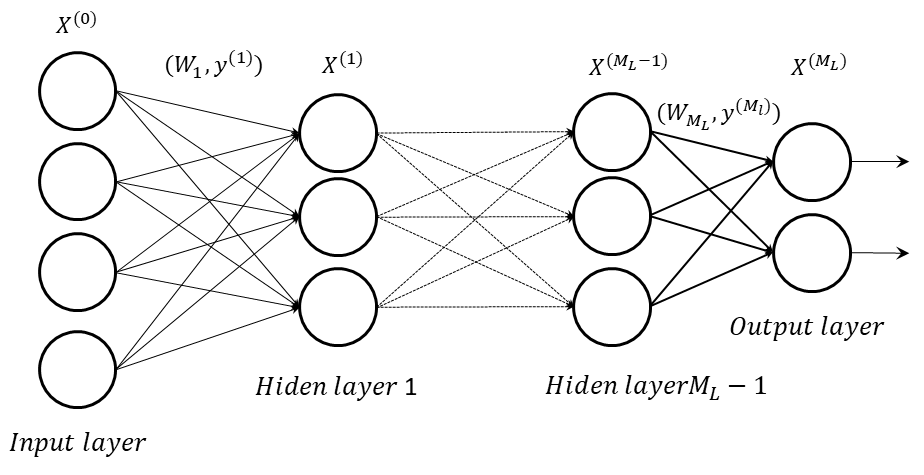}
\end{center}
\caption{\label{Fig3}Example of MultiLayer perceptron (MLP). }
\end{figure}
\subsection{Proposed secure multilayer perceptron}
  Securing a multilayer perceptron consists in implementing the feedforward and backward propagation phases over encrypted data. The MLP that we propose to secure, both in its learning and classification phase, is based on : i) perceptrons, the activation function of which ReLU, ii) the mean squared error (MSE)as cost function. The secure version of this MLP, we describe in the following, works with the Paillier cryptosystem and takes advantage of the above two servers based system architecture so as to exploit the secure multiplication and division operators depicted in Section \ref{section2}. As we will see, different issues have to be overcome so as to ensure the convergence of such a Secure MLP. In particular, we propose a new "Max" function operator so as to secure ReLU. 
\subsubsection{Secure MLP feed-forward phase}
The feedforward phase consists in calculating the MLP output for a given input. Based on the fact a MLP is constituted of different layers of perceptrons, securing the feed-forward phase relies on securing each perceptron independently. As seen in Section \ref{section2}, a perceptron primarily performs a weighted sum of the input vector $X=\{x_i\}_{i=1...n}$ (see eq. \eqref{weiSum}) the result of which is provided to an activation function (see eq. \eqref{relu}). Considering that all pieces of information provided by the user are Paillier encrypted, i.e. $\{E[x_i]\}_{i=1...n}$, the weighted sum in the encrypted domain becomes: 
	\begin{equation}
E[y]=\prod_{i=0}^n Mul_e^{P_1,P_2} (E[x_i],E[w_i])
\label{(13)}
\end{equation}
where $P_1$ and $P_2$ are the two independent servers (see Fig. \ref{Fig1}), $E[y]$ is the secure weighted sum and $\{E[w_i]\}_{i=1...n}$ the encrypted perceptron weights which are also confidential.

In this computation as well as in all others, one important constraint to consider stands on the fact that the Paillier cryptosystem only works with plain-text and cipher-text constituted of positive integers in $\mathbb{Z}_{K_p}$. More clearly, all data and parameters of the SMLP should be represented by integers. To overcome this issue, taking as example the input data, these ones are turned into integer values by scaling and quantizing them as follow
	\begin{equation}
X= [Qx]
\label{(12)}
\end{equation}
where $[.]$ is the rounding function and $Q$ is an expansion or scaling factor. Beyond, even if the SMLP parameters and inputs are integers, their processing may lead to negative values. In order to represent such values in $\mathbb{Z}_{K_p}$, integer values greater than $(K_p+1)/2$ will correspond to negative values and the others to positive values. 

By next, the secure perceptron's output is computed by applying a secure version of the ReLU activation function to the secure weighted sum $E[y]$. One key issue to overcome in securing ReLU (see eq. \eqref{relu}) stands in the calculation of the $Max(a,b)$ function in-between two integer values $a$ and $b$ in the Paillier encrypted domain. Different solutions have been proposed so as to securely compare encrypted data \cite{ding2017encrypted,wu2012secure,bellafqira2017secured,hsu2012image}. Most of them are based on blinding and two non-colluding parties. However, with all these approaches, the comparison result is provided in a clear form. More clearly, if $P_1$ asks $P_2$ to compare $E[a]$ and $E[b]$, $P_1$ will know if $E[a]$ is greater or not than $E[b]$. In our framework, this leads to an information leak. Indeed, $P_1$ is not authorized to get some information about the SMLP parameters. To solve this problem, and to the best of our knowledge, we propose a novel comparison operator $Comp_e^{P_1, P_2}$ the output of which is encrypted. It will be used so as to compute the $Max_e^{P_1, P_2}$ operator. $Comp_e^{P_1, P_2}$ works in two steps: 
  \begin{itemize}
  \item \textit{Data randomization} - $P_1$ selects two random values $r$ and $r'$ from $\mathbb{Z}_{K_p}$ such as $r'$ is significantly smaller than $r$ (i.e. $r>>r'$) and computes computes 
  \begin{equation}
E[r(a-b)-r']=(E[a]E[b]^{-1})^r\times E[r']^{-1}
\end{equation}	
 Then $P_1$ sends the result to $P_2$. 
  
 \item \textit{Secure comparison} -  $P_2$ decrypts the data and compare them to $0$ and sends an encrypted bit $i$ such that: 

\begin{equation}
   Comp_e^{P_1, P_2} (E[a],E[b])=E[i] =\left\{\begin{array}{l}
  E[1] \quad if\quad r(a-b)-r'>0 \\
  E[0] \quad if\quad $else$
\end{array}  \right.
\end{equation}	
Then $P_2$ sends $E[i]$ to $P_1$.

 Based on $Comp_e^{P_1, P_2} (E[a],E[b])$, $P_1$ can compute the $Max_e^{P_1, P_2}$ operator $Max_e^{P_1, P_2} (E[a],E[b])$, that is to say.
\begin{align*}
Max_e^{P_1, P_2} (E[a],E[b])&=E[max(a,b)]\\
&=Mul_e^{P_1, P_2} (E[a]E[b]^{-1},E[i])\times E[b]
\end{align*}
\begin{equation}
   =E[i(a-b)+b] =\left\{\begin{array}{l}
  E[a] \quad if\quad i=1 \\
  E[b] \quad if\quad i=0
\end{array}  \right.
\label{(11)}
\end{equation}	

As it can be seen, $P_1$ accesses to the encrypted version of the maximum value between two integers without knowing which value is greater than the other one. Such a security level is achieved based on the fact that Paillier cryptosystem uses random values which multiply after each multiplication (i.e. $E[a,r_1]E[b,r_2]=E[a+b,r_1 r_2]$ - see Section \ref{section2}). 
Finally, based on the $Max_e^{P_1, P_2}$ operator, the output of a secure ReLU based perceptron is given by: 
\begin{equation}
E[max(0,y)]=Max_e^{P_1, P_2} (E[0],E[y])
\end{equation}
\end{itemize}
To conclude, a secure MLP is based on secure perceptron layers. 

\subsubsection{Secure back-propagation phase} 
\label{section3-3}	
As stated in Section \ref{section3}, the objective of the back-propagation phase is to update the MLP weigths of each perceptron. In the supervised mode, for a given input, one computes the error between the MLP output and the input data label according to an objective or cost function. In this work the Mean Square Error (MSE) is used. Then MLP weights are updated so as to minimize this function.
 
Let us consider a MLP network composed of $M_L$ layers (see Fig. \ref{Fig3}) and an \textit{a priori} known vector input data $X^{(0)}$ along with its label $t$. If $X^{(M_L)}$ corresponds to the MLP output (see Fig. \ref{Fig3}), then the error $e$ in the clear domain is such as
\begin{equation}
e=MSE(X^{(M_L)},t)=||X^{(M_L)}-t||^2_2. 
\end{equation}
This cost function can be expressed in the Paillier encrypted domain under the following form
\begin{equation}
E[e]=Mul_e^{P_1, P_2} (E[X^{(M_L)}-t],E[X^{(M_L)}-t])
\end{equation} 
Let us recall, that in our framework, $P_1$ holds $E[X^{(M_L)}]$ and $E[t]$. It computes $E[X^{(M_L)}-t]$ thanks to the homomorphic Paillier proprieties, and interacts with $P_2$ so as to compute $E[e]=Mul_e^{P_1, P_2} (E[X^{(M_L)}-t],E[X^{(M_L)}-t])$ . 

Once the error computed, the next step stands in back propagating the error so as to update the MLP weights of each perceptron (i.e. $W=\{w_i\}_{i=1...n}$) so as to minimize the error value $e$. This update is based on the descent gradient. For sake of simplicity let us consider a MLP constituted of one single perceptron. The updated value $w_i^{'}$ of its weight $w_i$ is given by
\begin{align*}
w_i^{'} &= w_i  +\frac{1}{\lambda^{-1}} \frac{\partial e}{\partial w_i}\\
 		&= w_i+\frac{1}{\lambda^{-1}} 2(t-f(y))x_i f'(y)
\end{align*}
where $y=\sum_{i=1..n} w_i.x_i$, $\lambda$ is the learning rate factor, $f(.)$ is the activation function and $t$ the label of the data placed at the input of the MLP. Notice that, derivate of the ReLU function is $f'(y)=1_{(y>0)}$, where $1_{(.)}$ denotes the unit step function whose value is zero for negative input or one, otherwise. The same update operation in the encrypted domain, is 
\begin{equation}
E[w_i^{'}]=E[w_i]Div_e^{P_1, P_2}(E[\frac{\partial e}{\partial w_i}], \lambda)
\end{equation}
 with 
\begin{equation}
E[\frac{\partial e}{\partial w_i}]=Mul_e^{P_1, P_2} (E[t]^{-1} E[f(y)],E[x_i ]^2,E[1_{(y>0)}])
\end{equation}
	It is important to underline that the encrypted version of the unit step function $E[1_{(y>0)}]$ is equivalent to $E[1_{(y>0)}]=Comp_e^{P_1, P_2} (E[0],E[y])$. 

In the case of a MLP network of $M_L$ layers, with a set of labeled input data $T=\{X_i^{(0)},t_i\}_{i=1...K}$, where $K$ is the size of the training set and the error function for each input is 
\begin{equation}
e_{t_i}=||t_i-X_i^{(M_L)} ||^2
\end{equation}
		
Herein, $e_{t_i}$ depends on the weights of all MLP layers. The descent gradient can be computed with the help of the chain rule algorithm so as to calculate all partial derivatives, even those of intermediary layers. According to this algorithm, the last layer gradient is given as 
\begin{equation}
\frac{\partial e_{t_i}}{\partial W_i^{(M_L)}} =\delta^{(M_L)} .^{tr} X_i^{(M_L-1)}=(X_i^{(M_L)} -t_i)*f'(W_i^{(M_L)}X_i^{(M_L-1)}) .^{tr}X_i^{(M_L-1)}
\end{equation}
where the operator $*$ denotes the product element wise (aka Hadamard product). The gradient computation for any hidden layer $l$ is recursively defined as 
\begin{align*}
	\frac{\partial e_{t_i}}{\partial W_i^{(l)}}& =  \delta_i^{(l)} .^{tr}X_i^{(l-1)}\\
	&=W_i^{(l+1)} \delta_i^{(l+1)}* f'(W_i^{(l)}X_i^{(l-1)}).^{tr}X_i^{(l-1)}
\end{align*}
where $tr$ indicate the transpose vector.
 
As a consequence the update of each layer weight is given as 
\begin{equation}
	W_i^{(l)}=W_i^{(l)}+\frac{1}{\lambda^{-1}} \frac{\partial e_{t_i}}{\partial W_i^{(l)}}
\end{equation}	
where $\lambda$ is the global learning rate factor. 
The back-propagation phase in the encrypted domains can be easily derived, and the update version of a a perceptron layer $l$ is given by 
\begin{align*}
	E[\frac{\partial e_{t_i}}{\partial W_i^{(l)}}]& =  Mul_e^{P_1,P_2}(E[\delta_i^{(l+1)}], E[.^{tr} X_i^{(l-1)}])\\
	&=Mul_e^{P_1,P_2}(E[W_i^{(l+1)}], E[\delta_i^{(l+1)}]*_e E[f'(W_i^{(l)}], E[X_i^{(l-1)})], E[.^{tr}X_i^{(l-1)}])
\end{align*}
where the $*_e$ operator is the secure version of the element wise multiplication $*$ defined as for two vectors of the same size $n$ 
\begin{equation}
E[u]*E[v]=(Mul_e^{P_1, P_2}(E[u_1], E[v_1]),...,Mul_e^{P_1, P_2}(E[u_n], E[v_n]))
\end{equation}
By iteratively applying the secure feedforward and back-propagation phases, it is possible to train our Secure MLP without compromising the security of the input data and of the SMLP parameters. It is the same when classifying new data. The services class providers will have no idea of the MLP output. 
 
 \begin{figure}[h!]
\begin{center}
\includegraphics[scale=.4]{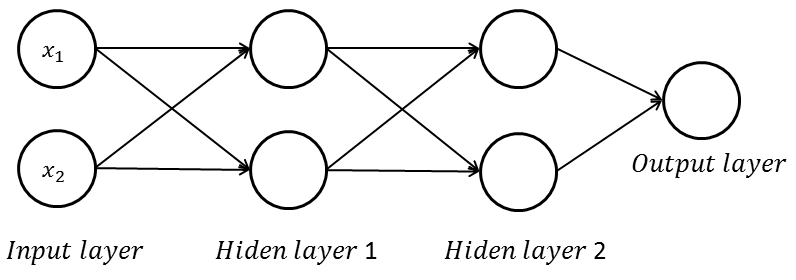}
\end{center}
\caption{\label{Fig4} SMLP architecture used to learning the AND function}
\end{figure}

\section{Experimental results and security analysis}
 \label{section4}
 The proposed SMLP solution has been implemented so as to learn the $And-Logic$ function in a supervised training mode. This function takes as input two real numbers $x_1$ and $x_2$ in the interval $[0,1]$ and its output is a binary value such that: 
\begin{equation}
 y=[x_1]AND[x_2]
 \label{ch4:eqET}
 \end{equation}
where $[.]$ denotes the rounding operator. 
\subsection{Dataset and MLP architecture}
The training data set is constituted of 10000 lines of three columns each, where each line represents a training input sample. The first two columns contain two real values between 0.0 and 1.0, while the last column contains their $AND$ value. 

Fig. \ref{Fig4} provides the architecture of the implemented SMLP. It is composed of an input layer and of two hidden layers, both containing two perceptrons, followed by an output layer of one perceptron. As stated above, the network is based on our secure ReLU activation function (see section \ref{section3}). In all following test, the expansion factor $Q$ was fixed to $10^6$ so as to ensure the SMLP works with integer values with a training phase limited to 100 epochs 2000 samples of the training data set are used for the learning phase and the 8000 other for the testing phase. The expected result is that, upon the entry of two values contained between 0.0 and 1.0, the activation of the output layer after feedforward contains the value of the $AND$ between the two inputs, that is, either 0 or 1.

\subsection{Secure MLP performance}
\label{section4-1-2}
The performance of our secure MLP, which is expressed in terms of classification accuracy and also convergence, depends on the learning rate. Precision is the number of correct predictions made divided by the total number of predictions made. We recall that the learning rate factor $\lambda$ plays a critical role in the convergence of the network. Indeed, it influences the modification step in the weight update in the back-propagation phase (see Section \ref{section3}). We tested several $\lambda$ values in the range $[10^{-12};10^{-4}]$. We give in Table \ref{ch4:tabtaux}, the precision of our SMLP in average after $10$ tries, and if yes or no it has converged after the training. It can be seen that SMLP converges for values of $\lambda$ smaller than $\lambda =10^{-8}$. We thus recommend taking initial weights distributed in the range $[10^{-5},10^5]$ and a learning rate factor $\lambda =10^{-10}$.

\begin{table}
\caption{Convergence and precision of our SMLP for different learning rate factor values. }
\centering 
\begin{tabular}{|c|c|c|c|c|c|}
\hline 
$\lambda$ & $10^{-12}$ & $10^{-10}$ & $10^{-8}$ & $10^{-6}$ & $10^{-4}$ \\ 
\hline 
Convergence (Y/N)/Accuracy & Yes/$72\%$ & Yes/$83\%$ & Yes/$93.1\%$  & No & No \\ [.1cm]
\hline 
\end{tabular} 
\label{ch4:tabtaux}
\end{table}
We also have trained the equivalent MLP in the clear domain under the same conditions  with a learning rate factor of 0.005 and 100 epochs. The obtained MLP precision is about $98.3\%$ with of course no convergence issues. It can be seen based on Table \ref{ch4:tabtaux}, that SMLP always provides lower performance. This can be explained by the use of an expansion factor so as to convert real values into integer values. Anyway, these results demonstrate it is possible to train a MLP in a secure outsourced way.

\subsection{Security analysis under the semi-honest model}
\label{section4-2}
The following analysis considers the semi-honest cloud adversary model as presented in Section \ref{section3}. Due to the fact all data (i.e. input data and SMLP parameters) are encrypted with the Paillier cryptosystem the security of which has been demonstrated in \cite{paillier1999public}, the security of the feed-forward and back-propagation phases stands on the security of the operators $Mul_e^{P_1, P_2} (.)$, $Div_e^{P_1, P_2} (.)$ and  $Comp_e^{P_1, P_2} (.)$.
\begin{itemize}
\item Security of $Mul_e^{P_1, P_2}(.)$ -  As shown in Section \ref{section2}, $Mul_e^{P_1, P_2} (E[a],E[b])$ relies on a data blinding operation $P_1$ applies on $E[a]$ and $E[b]$ to compute $E[a\times b]$. To do so, $P_1$ generates two random values $r_a$ and $r_b$ from $\mathbb{Z}_{K_p}$ and computes $E[a+r_a]$ and $E[b+r_b]$. $P_2$ decrypts by next these values. Since $r_a$ and $r_b$ are randomly chosen in $\mathbb{Z}_{K_p }$ and only known from $P_1$, they give no clues to $P_2$ regarding $a$ and $b$. 
  \item Security of $Div_e^{P_1, P_2} (.)$ -  We let the reader refer to \cite{veugen2010encrypted}, where Thijs Veugen proved the security of the operator $Div_e^{P_1, P_2} (.)$.
  \item Security of $Comp_e^{P_1, P_2} (.)$ - and consequently of $Max_e^{P_1, P_2} (.,.)$ 
As explained in Section \ref{section3}, $Max_e^{P_1, P_2}(E[a],E[b])=E[max(a,b)]$ depends on the $Comp_e^{P_1, P_2}(;)$ operator. Let us consider $P_1$ possesses the couple $(E[a],E[b])$ and that he wants to compute  $Max_e^{P_1, P_2} (E[a],E[b])$. To do so, it computes $E[r(a-b)-r']$ where $r$ and $r'$ are chosen uniformly from $\mathbb{Z}_{K_p}$ under the constraint that $r>>r'$. $P_2$ accesses to $r(a-b)-r'$ from which it cannot deduce any information about $a$ and $b$ nor about $a-b$ since it does not know $r$ and $r'$. $P_2$ compares this value to zero. This comparison gives not more information to $P_2$. 

By next, in order to avoid that $P_1$ knows the comparison result, $P_2$ encrypts using the user public key the bit 0 or 1 (see Section  \ref{section3}) it sends to $P_1$. $P_1$ can derive the results of the function $Max_e^{P_1, P_2} (E[a],E[b])$, because all these computation are conducted over encrypted data, $P_1$ has no idea about $a$,  $b$ and $Max(a,b)$. 

The rest of the computations (e.g. MSE, error derivatives) are based on either encrypted or randomized data. As consequence, if $P_1$ and $P_2$ do not collude, no information related to the user data or to the SMLP model is disclosed. Since all operations involved in the computation of the feed-forward and back-propagation phases are in cascade then, according to the sequential Composition theorem \cite{goldreich2009foundations}, SMLP is completely secure under the semi-honest model. 
\end{itemize}
  
 \section{Conclusion}
 \label{section5}
 In this paper, we have proposed a new Secure Multilayer Perceptron (SMLP) which can be deployed in the cloud. Its main originality, compared to actual homomorphic encryption based SMLP schemes, is that it can be trained with homomorphically encrypted data with no extra communications between the user and the servers. With this scheme, all data, input data, SMLP output and its parameters, are encrypted. Our SMLP is based on: an original secure version of the $Max(.,.)$ function we propose, the result of which is encrypted; a ReLU activation function secured with no linear approximation. Such a SMLP has been implemented so as to learn or model the AND function in-between real values. Experimental results demonstrate that SMLP converges in its training phase under some parameter initialization constraints. Beyond the complexity of our SMLP, which is based on homomorphic encryption, these preliminary results are very encouraging.  

 \bibliographystyle{splncs04}
 \bibliography{biblio}
%

\end{document}